\documentclass[twoside]{ae100prg}
\usepackage{graphicx}
\usepackage[breaklinks]{hyperref}
\usepackage{booktabs}
\usepackage{amsmath,amssymb,amsfonts} % die mathe schrifent

\newcommand{\pd}[2]{\frac{\partial #1}{\partial #2}}

\begin{document}
\title{Coupling  Dimers to CDT}

\author{L. Glaser }

\address{The Niels Bohr Institute, Copenhagen University,\\
Blegdamsvej 17, DK-2100 Copenhagen \O , Denmark }

%\address{$^b$~Institute of Physics, Jagiellonian University,\\
%Reymonta 4, PL 30-059 Krakow, Poland.\\
%email: atg@th.if.uj.edu.pl\\}

%\address{$^c$~Department of Physics, Nagoya University, \\
%Nagoya 464-8602, Japan.\\
%email: ysato@th.phys.nagoya-u.ac.jp}

\email{glaser@nbi.dk}

\begin{abstract}
This contribution reviews some recent results on dimers coupled to CDT. A bijective mapping between dimers and tree-like graphs allows for a simple way to introduce dimers to CDT. This can be generalized further to obtain different multicritcal points.
\end{abstract}

\section{Introduction}
Causal Dynamical Triangulations (CDT) is a proposed theory of quantum gravity. In CDT the path integral for gravity is regularized through simplices as in dynamical triangulation. CDT introduces a preferred time slicing to provide for a well-defined Wick rotation. This preferred time slicing leads to a better behaved continuum theory \cite{ambjorn98} (see \cite{ambjorn_nonperturbative_2012} for a review).

For matrix models it is well-know that random lattices can be coupled to matter, like dimers or the Ising model, to find quantum gravity coupled to conformal field theories \cite{kazakov_appearance_1989,matthias_yang-lee_1990}. It is then an interesting prospect to try and couple matter to the random lattices of CDT.

In this article we review the results obtained in \cite{ambjorn_new_2012} and present a simple extension of the model which allows for higher order multicritical points.\footnote{ Quite similar results have been obtained simultaneously in \cite{Atkin:2012yt,Atkin:2012ka}}
\section{Coupling CDT to Dimers }
\begin{figure}
\centering
\begin{minipage}{.45\textwidth}
  \centering
\includegraphics[width=0.55\linewidth]{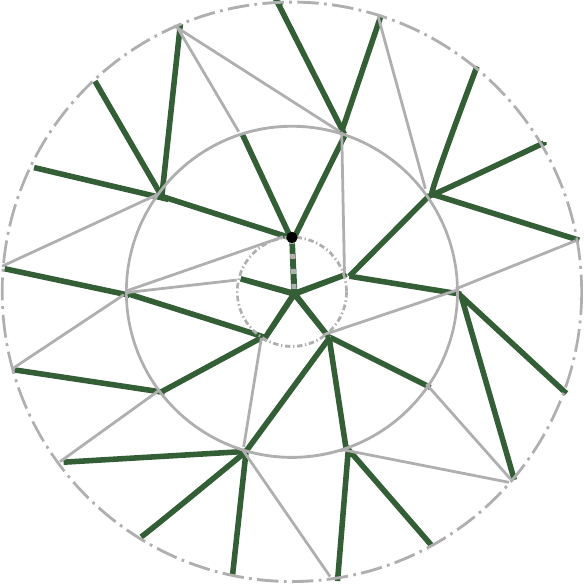}
\caption{\label{fig:bijection} The green marked links are those parts of the CDT that are part of the graph. If only these were present we could still reproduce the CDT by reintroducing the space-like links and the leftmost link at every vertex.}
\end{minipage}\hfill
\begin{minipage}{.45\textwidth}
  \centering
\includegraphics[width=0.9\linewidth]{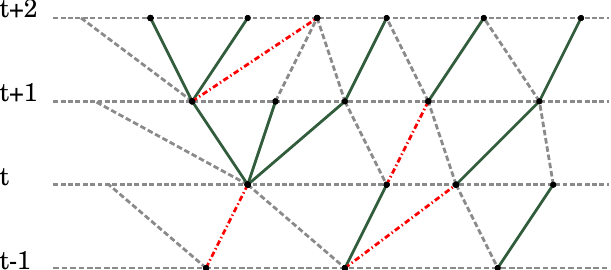}
\caption{\label{fig:dimers}This is a cut-out of some CDT with the tree graph marked green. Dimers are highlighted in red, and can only be parts of the tree.}
\end{minipage}
\end{figure}

Durhuus et al.\cite{durhuus09} proved that there is a bijective mapping between rooted tree graphs \ref{fig:bijection} and CDT.
This bijection makes it possible to determine the critical exponents of CDT using recursive equations as in \cite{QuGeom}. It also makes it possible to consider the easier problem of coupling dimers to a rooted tree graph instead of directly placing them on the CDT. The simple rule of placing any number of hard  dimers on the tree will lead to a partition function which allows for new multicritical behavior\cite{ambjorn_new_2012}.

\begin{figure}
\begin{minipage}{.45\textwidth}
  \centering
\includegraphics[width=0.85\textwidth]{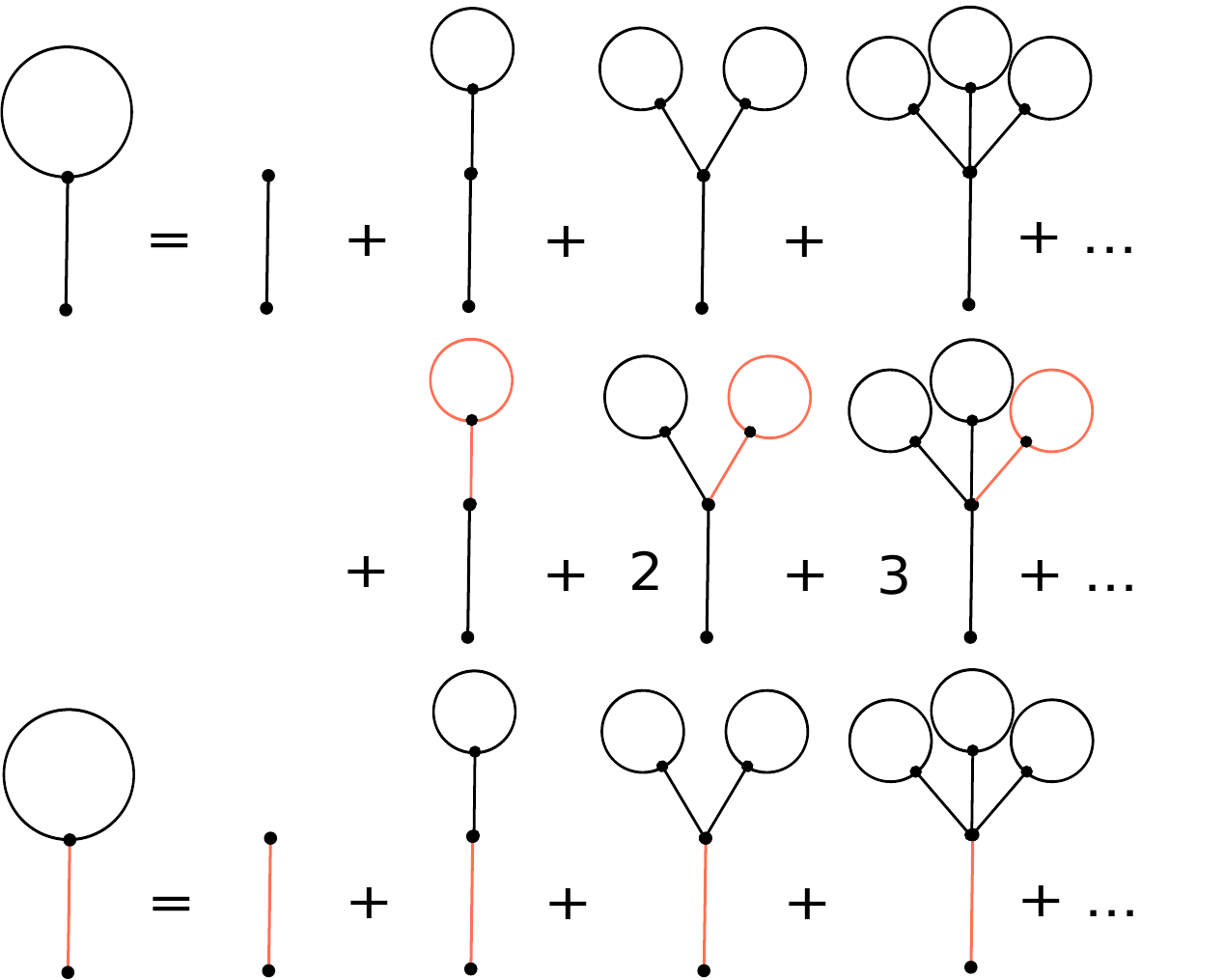}
\caption{\label{fig:recursion} The recursive equation for the branched polymers. The circle stands for all possible configurations.}
\end{minipage}\hfill
\begin{minipage}{.45\textwidth}
  \centering
  \includegraphics[width=0.85 \textwidth]{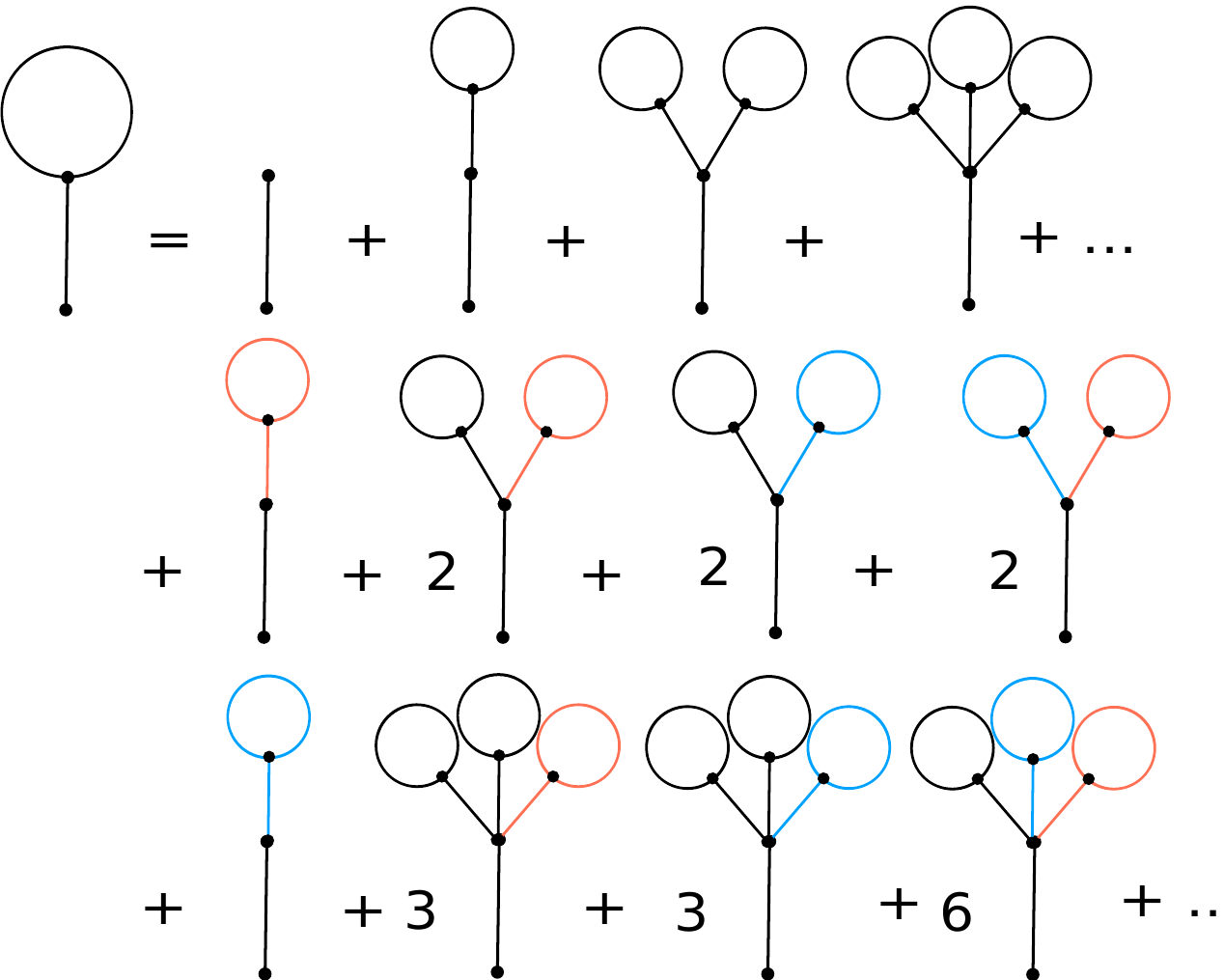}
\caption{\label{fig:recursion_higherorder} For a higher order multicritcal point there are two types of dimers (red and blue). The recursive equation starting in a dimer is the same as in figure \ref{fig:recursion}.}
\end{minipage}
\end{figure}

The partition function for CDT with dimers reads
\begin{equation}
Z(\mu ,\xi ) = \sum_{BP}  e^{-\mu } \sum_{{\rm HD(BP)}} \xi^{|{\rm HD(BP)}|} \quad.
\end{equation}
where $\rm BP$ is the set of all tree-like graphs, $\rm HD(BP)$ the set of all dimer configurations on that graph and $\rm HD(BP)$ the number of dimers in a given configuration. This partition function can be solved using recursive equations which arise for the tree like graphs and are discussed in detail in \cite{QuGeom}.
The recursion depicted in figure \ref{fig:recursion} leads to the equations 
\begin{align}\label{eq:recursepartitions}
Z&= e^{-\mu} \left( \frac{1}{1-Z} +  W \frac{1}{(1-Z)^2} \right) &	W&= e^{-\mu}\xi \left( \frac{1}{1-Z} \right) \quad,
\end{align}
where $Z$ is the partition function for a tree with a normal link at the root and $W$ is the partition function for a tree rooted in a dimer.
At a $n$-multicritical point the first $n-1$ derivatives of the coupling $\mu$ by the partition function $Z$ are zero
\begin{equation}
\pd{\mu}{Z}\bigg|_{Z_c}= \dots = \pd{^{n-1} \mu}{Z^{n-1}} \bigg|_{Z_c}=0 \;.
\end{equation}
We can then solve equations \eqref{eq:recursepartitions} to find the third multicritical point at
\begin{equation}
		Z_c=   \frac{5}{8}  \hspace{20pt}		\xi_c= 	-\frac{1}{12} \hspace{20pt}			  e^{\mu_c}=  \frac{32}{9} \quad ,
\end{equation}
with the critical exponents being 
\begin{equation}
\gamma =\frac{1}{3} \qquad d_H=\frac{3}{2} \qquad \sigma= \frac{1}{2} \;.
\end{equation}
In pure CDT, not coupled to dimers, one finds
\begin{equation}
\gamma= \frac{1}{2} \qquad d_H=2 \qquad \sigma= \text{not defined}
\end{equation}
so it is clear that CDT coupled to dimers lies in a different universality class than pure CDT and therefore represents an interacting system of matter and gravity. However the negative weight $\xi$ does make the physical interpretation of the results less clear \cite{glaser_MG13_2012}.
%\section{Higher order multicritical point through inclusion of different dimers}
It is easy to generalize this model to higher order multicritcal points. To do so one introduces different types of dimers, denoted as type $a$ with weight $\xi$ and type $b$ with weight $\zeta$.
The rule is then that a vertex with an incoming dimer can not spawn any type of dimer, while a vertex with an incoming empty link can spawn at most one dimer of each color. This is illustrated in figure \ref{fig:recursion_higherorder}.
This model leads to the partition function
\begin{equation}
Z(\mu ,\xi ,\zeta ) = \sum_{\rm BP} e^{-\mu } \sum_{{\rm HD(BP(a,b))}} \xi^{|a|} \zeta^{|b|} \quad .
\end{equation}
where $\rm HD(BP(a,b))$ denotes the set of configurations of hard dimers of type $a$ and $b$ and $|a|$ (resp $|b|$) is the number of dimers of type $a$ (resp $b$) in the configuration. 
It can again be solved using recursive equations for the tree graphs
\begin{align}
Z= e^{- \mu} \left(\frac{1}{1-Z} + \frac{1}{(1-Z)^2} (W+V) + \frac{W V}{(1-Z)^3} \right) \quad ,
\end{align}
where $W$ denotes the partition function starting in a dimer of type $a$  and $V$ for dimers of type $b$. For $W$ and $V$ we obtain equations like in \eqref{eq:recursepartitions}.% (for $V$ one replaces $\xi$ by $\zeta$).

This model has one multicritical point of fourth order at $(\xi_c,\zeta_c)=\left (\frac{1}{90} \left(5 \mp i \sqrt{35}\right),\frac{1}{90} \left(-5 \pm i \sqrt{35}\right) \right)$ and $e^{-\mu_c} =\frac{256}{75}$. 
%For this model the weights are not only negative, as before but now they are complex which makes the physical interpretation even more challenging. 
The critical exponents are
\begin{align}
\gamma&=\frac{1}{4} & d_{H}&=\frac{4}{3} \quad.
\end{align}
It is possible to extend this model to any further multicritical point by introducing additional colors of dimers. % obeying the same rules as for two different colors.
\section{Summary}
Introducing dimer-like matter to CDT leads to new critical behavior. This means that there is a coupling between the quantum gravity of CDT and the matter of the dimers. 
Through the introduction of different types of dimers it is possible to obtain multicritical points of any order.
\section*{Acknowledgments}
I would like to thank the Danish Research Council for financial support via the grant ``Quantum gravity and the role of Black holes''. 
\section*{References}
\bibliography{bibliography}

\end{document}